\documentclass[a4paper,10pt,eqno]{article}
\usepackage{theorem}
\usepackage{latexsym,amssymb,amsfonts,amsmath}
\usepackage{cite}
\newtheorem{thm}{Theorem}[section]
\newtheorem{lem}[thm]{Lemma}

\theoremheaderfont{\scshape}

\newcommand{\EM}{\mathbb{E}}
\newcommand{\PM}{\mathbb{P}}

\newcommand{\qed}{\hfill $\Box$}

\newcommand{\ket}[1]{|#1\rangle}
\newcommand{\bra}[1]{\langle#1|}

\title{{\Large {\bf Time averaged distribution of a discrete-time quantum walk\\ on the path}}}
\author{
{\small Yusuke Ide\footnote{To whom correspondence should be addressed. E-mail: ide@kanagawa-u.ac.jp}}\\
{\scriptsize Department of Information Systems Creation, 
Faculty of Engineering, 
Kanagawa University}\\
{\scriptsize Kanagawa, Yokohama 221-8686, Japan}\\
{\scriptsize e-mail: ide@kanagawa-u.ac.jp, Tel.: +81-45-481-5661, Fax: +81-45-413-6565}\\
{\small Norio Konno}\\
{\scriptsize Department of Applied Mathematics, 
Faculty of Engineering, 
Yokohama National University}\\
{\scriptsize Hodogaya, Yokohama 240-8501, Japan}\\
{\scriptsize e-mail: konno@ynu.ac.jp, Tel.: +81-45-339-4205, Fax: +81-45-339-4205}\\
{\small Etsuo Segawa}\\
{\scriptsize Graduate School of Information Science, 
Tohoku University}\\
{\scriptsize Aoba, Sendai 980-8579, Japan}\\
{\scriptsize e-mail: e-segawa@m.tohoku.ac.jp, Tel.: +81-22-795-3218, Fax: +81-22-795-3218}\\
}
\date{\empty }
\begin{document}
\maketitle

\par\noindent
\begin{small}
\par\noindent
{\bf Abstract}
\newline 
The discrete time quantum walk which is a quantum counterpart of random walk plays important roles in the theory of quantum information theory. In the present paper, we focus on discrete time quantum walks viewed as quantization of random walks on the path. We obtain a weak limit theorem for the time averaged distribution of our quantum walks. 
\footnote[0]{
{\it Abbr. title:} Time averaged distribution of a DTQW on the path
}
\footnote[0]{
{\it AMS 2000 subject classifications: }
60F05, 60G50, 82B41, 81Q99
}
\footnote[0]{
{\it PACS: } 
03.67.Lx, 05.40.Fb, 02.50.Cw
}
\footnote[0]{
{\it Keywords: } 
Quantum walk, Path, Jacobi matrix
}
\end{small}

\setcounter{equation}{0}
\section{Introduction}
In the last decade, the discrete-time quantum walk (DTQW) has been extensively studied by many authors as a quantum counterpart of the random walk which plays important roles in various fields \cite{Kempe2003,Kendon2007,VAndraca2008,Konno2008b,AharonovEtAl2001,AmbainisEtAl2001,AhlbrechtEtAl2011}. In the theory of quantum algorithm, quantum walk also plays important roles. In fact, quantum search algorithms are often reduced to quantum walks on the path, for example, search algorithms on the hypercube \cite{ShenviEtAl2003} or glued binary tree \cite{ChildsEtAl2003} and an algorithm for element distinctness on the Johnson graph \cite{Ambainis2004}. Therefore, investigations of DTQW on the path is beneficial. 

Rohde {\em et al}.\ \cite{RohdeEtAl2011} studied periodic properties of entanglement for DTQW on the path determined by biased Hadamard coins numerically. Recently, Godsil \cite{Godsil2011} studied the time averaged distributions of continuous-time quantum walks (CTQW) on the path using the average mixing matrix. In the case of CTQW on the path determined by the graph Laplacian, the time averaged distribution consists of a uniform distribution and additional masses on the starting vertex and a vertex located at the symmetrical position about the middle point of the graph. In the present paper, we investigate the time averaged distribution of a DTQW. The quantum walk treated here can be viewed as a quantization of a random walk on the path. We should remark that the construction of our DTQW is closely related to that of Szegedy's work \cite{Szegedy2004}. 

The rest of this paper is organized as follows. The definition of our DTQW is given in Sect.\ 2 and main result of this paper is stated in Sect.\ 3. The remaining section (Sect.\ 4) is devoted to the proof of our result. 
\section{Definition of the DTQW}
In this paper, we consider a discrete-time quantum walk on the path $P_{n+2}$ with $n+2$ numbers of vertices $V_{n+2}= \{0,1,\ldots ,n,n+1\}$ and the edges $E_{n+2}= \{(i,i+1):i=0,1,\ldots ,n\}$. For sake of simplicity of expressions, we choose $P_{n+2}$ because the number of the intermediate vertices in the graph equals $n$. From now on we use $n$ for suffix of operators although the number of the vertices of the graph equals $n+2$. In order to define the DTQW, we use a Hilbert space $\mathcal{H}_{n}= \mathrm{span}\{\ket{0,R}, \ket{1,L}, \ket{1,R},\ldots ,\ket{n,L},\ket{n,R},\ket{n+1,L}\}$ with $\ket{i,J}=\ket{i}\otimes \ket{J}\ (i\in V_{n+2}, J\in \{L,R\})$ the tensor product of elements of two orthonormal bases $\{\ket{i}:i\in V_{n+2}\}$ for position of the walker and $\{\ket{L}={}^T [1,0], \ket{R}={}^T [0,1]\}$ for the chirality which means the direction of the motion of the walker where ${}^T \!\!A$ denotes the transpose of a matrix $A$. Then we consider the time evolution operator $U^{(n)}$ on $\mathcal{H}_{n}$ defined by $U^{(n)}=S^{(n)}C^{(n)}$ with the coin operator $C^{(n)}$ and the shift operator $S^{(n)}$ defined as follows:
\begin{align*}
C^{(n)}&=\ket{0}\bra{0}\otimes \ket{R}\bra{R}+\sum_{j=1}^{n}\ket{j}\bra{j}\otimes \left(2\ket{\phi}\bra{\phi}-I_{2}\right)+\ket{n+1}\bra{n+1}\otimes \ket{L}\bra{L},\\
S^{(n)}\ket{i,J}&=
\begin{cases}
\ket{i+1,L}&\text{if}\ \ J=R,\\ 
\ket{i-1,R}&\text{if}\ \ J=L,
\end{cases}
\end{align*}
where $\ket{\phi}=\sqrt{q}\ket{L}+\sqrt{p}\ket{R}\ (0<p,q<1,\ p+q=1)$ and $I_{n}$ be the $n\times n$ identitiy matrix. 

Let $X_{t}^{(n)}$ be the position of our quantum walker at time $t$. The probability that the walker with initial state $\ket{\psi}$ is found at time $t$ and the position $x$ is defined by 
\begin{eqnarray*}
\PM_{\ket{\psi}}(X_{t}^{(n)}=x)=\left\lVert \left(\bra{x}\otimes I_{2}\right)\left(U^{(n)}\right)^{t}\ket{\psi}\right\rVert^{2}.
\end{eqnarray*}
In this paper, we consider the DTQW starting from a vertex $i\in V_{n}$ and choose the initial chirality state with equal probability, i.e, we choose the initial state as $\ket{\psi}_{0}=\ket{0}\otimes \ket{R}$ for $i=0$, $\ket{\psi}_{i}=\ket{i}\otimes \ket{L}$ or $\ket{\psi}_{i}=\ket{i}\otimes \ket{R}$ with probability $1/2$ for $1\leq i\leq n$, or $\ket{\psi}_{n+1}=\ket{n+1}\otimes \ket{L}$ for $i=n+1$. For the sake of simplicity, we write $\PM_{i}(X_{t}^{(n)}=x)$ for $\PM_{\ket{\psi}_{i}}(X_{t}^{(n)}=x)$. We consider the time averaged distribution 
\begin{eqnarray*}
\bar{p}_{i}^{(n)}(x)=\lim _{T\to \infty }\EM\left[\frac{1}{T}\sum_{t=0}^{T-1}\PM_{i}(X_{t}^{(n)}=x)\right],
\end{eqnarray*} 
where the expectation takes for the choice of the initial chirality state.
\section{Main result}
Let $\overline{X}_{i}^{(n)}$ be a random variable with distribution $\overline{p}_{i}^{(n)}$, i.e., $\PM (\overline{X}_{i}^{(n)}=x)=\overline{p}_{i}^{(n)}(x)$, we have the following weak limit theorem for $\overline{X}_{i}^{(n)}$ scaled by the system size $n$: 
\begin{thm}\label{thmtimeavedist}
\begin{enumerate}
\item 
If the starting vertex $i$ is fixed (finite) then 
\begin{eqnarray*}
\frac{\overline{X}_{i}^{(n)}}{n}\Rightarrow c_{i}\cdot \delta _{0}+(1-c_{i})\cdot U(0,1)\quad (n\to \infty),
\end{eqnarray*}
where $U(0,1)$ and $\delta _{0}$ are the uniform distribution on $[0,1]$ and the delta measure at $0$, respectively, and 
\begin{eqnarray*}
c_{i}=
\begin{cases}
I_{\{p<q\}}\cdot \left(1-\frac{p}{q}\right)& \text{if $i=0$},\\
I_{\{p<q\}}\cdot \frac{1}{2q}\left(1-\frac{p}{q}\right)\left(\frac{p}{q}\right)^{i-1}& \text{if $1\leq i<\infty $},
\end{cases}
\end{eqnarray*}
with the indicator function $I_{A}$, i.e., $I_{A}=1$, when $A$ is true and $I_{A}=0$, otherwise. Here $\Rightarrow $ represents the weak convergence. 
\item 
Let $f(n)$ be a function such that $rf(n)\leq n$ with $0<r<1$ and 
$f(n)\to \infty $ as $n\to \infty $. 
If the starting vertex $i$ is satisfing $i/f(n)\to r$ as $n\to \infty $ then
\begin{eqnarray*}
\frac{\overline{X}_{i}^{(n)}}{n}\Rightarrow U(0,1)\quad (n\to \infty).
\end{eqnarray*}
\end{enumerate}
\end{thm}
The idea of the proof of Theorem \ref{thmtimeavedist} is based on \cite{Szegedy2004}. In this proof, the eigenspace of the following $(n+2)\times (n+2)$ finite Jacobi matrix $J_{n+2}$ induced by a random walk on $P_{n+2}$ which is reflected with probability $1$ at the boundaries and moves to the right and the left with probability $p$ and $q$ respectively, plays an important role: 
\begin{eqnarray*}
J_{n+2}=
\begin{bmatrix}
0 & \sqrt{q} & & & & & \\
\sqrt{q} & 0 & \sqrt{pq} & & & \mbox{\smash{\huge\textit{O}}} & \\
 & \sqrt{pq} & \ddots & \ddots & & & \\
 & & \ddots & \ddots & \sqrt{pq} & \\
 & & & \sqrt{pq} & 0 & \sqrt{p}\\
\mbox{\smash{\huge\textit{O}}} & & & & \sqrt{p} & 0
\end{bmatrix}. 
\end{eqnarray*}
In addition, $c_{i}$ in Theorem \ref{thmtimeavedist} is closely related to the stationary measure of the corresponding random walk. Let $\pi_{n}(i)\ (i=0,\ldots ,n+1)$ be the stationary measure of the corresponding random walk on $P_{n+2}$. By simple calculations, we have
\begin{eqnarray*}
\pi_{n}(i)=
\begin{cases}
\frac{1-p/q}{2\left\{1-(p/q)^{n+1}\right\}} &\text{if $i=0$,}\\
\frac{1-p/q}{2q\left\{1-(p/q)^{n+1}\right\}}(p/q)^{i-1} &\text{if $i=1,\ldots ,n$},\\
\frac{1-p/q}{2\left\{1-(p/q)^{n+1}\right\}}(p/q)^{n} &\text{if $i=n+1$},
\end{cases}
\end{eqnarray*}
for $p\neq q$ and 
\begin{eqnarray*}
\pi_{n}(i)=
\begin{cases}
\frac{1}{2(n+1)} &\text{if $i=0, n+1$,}\\
\frac{1}{n+1} &\text{if $i=1,\ldots ,n$},
\end{cases}
\end{eqnarray*}
for $p=q=1/2$. This suggests that $c_{0}/2=\lim _{n\to \infty }\pi_{n}(0)$ and $c_{i}=\lim _{n\to \infty }\pi_{n}(i)\ (1\leq i<\infty )$. In fact, $\pi_{n}(i)$ is the same as square
of the $i$-th component of the eigenvectors corresponding to the eigenvalue $1$ and $-1$ 
of $J_{n+2}$. These eigenvectors give the mass at the origin.

It is well-known that the ballistic spreading of both CTQW and DTQW on the 
``infinite'' line are characterized by weak limit theorems whose limit 
densities are quite similar except the extra term $(1-x^{2})^{-1}$ in 
DTQW case \cite{Konno2008b}. We should remark that a relation between CTQW and 
DTQW on the infinite line is discussed in \cite{Strauch2006}. We also see 
more detailed discussion on the relations with respect to its weak limit theorems 
in \cite{ChisakiEtAl2011}. 
In the case of CTQW determined by graph Laplacian on $P_{n+2}$ \cite{Godsil2011}, the time averaged distribution is almost the same as the uniform distribution. If we start from a vertex $i\in V_{n+2}$, additional mass $1/(2n+2)$ is added to only two vertices $i$ and $n+1-i$. Because the additional masses tend to zero as $n\to \infty$, the limit distribution of CTQW corresponding to Theorem \ref{thmtimeavedist} is the uniform distribution $U(0,1)$. Therefore, the localization at the origin in the sense of Theorem \ref{thmtimeavedist} is not observed in the case of the CTQW. On the other hand, we can see in \cite{ChisakiEtAl2010} a weak limit theorem for a DTQW $X_{t}^{(\infty )}$ on the half line starting from the origin whose system is infinite in advance. In this case, 
\begin{eqnarray*}\label{thmCKS}
\frac{X_{t}^{(\infty )}}{t}\Rightarrow c_{0}\cdot \delta _{0}+(1-c_{0})\cdot K(2\sqrt{pq})\quad (t\to \infty),
\end{eqnarray*}
where $K(r)\ (0<r<1)$ has the following density:
\begin{eqnarray*}
f(x;r)=\frac{x^{2}}{1-\sqrt{1-r^{2}}}\cdot f_{K}(x;r)\cdot I_{[0,\infty ]}(x),
\end{eqnarray*}
where $f_{K}(x;r)$ is the Konno density function \cite{Konno2002, Konno2005}. This suggests that infinite nature of underlying graph changes the shape of limit distribution from a uniform distribution. 
It is a future problem finding connection between CTQW and DTQW on finite 
graphs in the sense of \cite{Strauch2006}.  
\section{Proof of Theorem \ref{thmtimeavedist}}
In order to prove Theorem \ref{thmtimeavedist}, we consider the eigenspace of the Jacobi matrix $J_{n+2}$ at first. 
\begin{lem}\label{eigenJn}
\begin{enumerate}
\item 
Let $\lambda _{m}\ (m=0,1,\ldots ,n,n+1)$ be the eigenvalues of the matrix $J_{n+2}$. Then we have 
\begin{align*}
\lambda _{0}&=1,\\
\lambda _{m}&=2\sqrt{pq}\cos \theta _{m}\ (m=1,\ldots ,n),\\
\lambda _{n+1}&=-1, 
\end{align*}
where $\theta _{m}=m\pi/(n+1)$. 
\item 
Let $\tilde{\mathbf{v}}_{m}$ be the eigenvector corresponding to an eigenvalue $\lambda _{m}$ for each $0\leq m\leq n+1$. Then we have 
\begin{align*}
{}^T \tilde{\mathbf{v}}_{0}&=\sqrt{\frac{p-q}{\left(p/q\right)^{n+1}-2q}}\cdot \left[1,\frac{1}{\sqrt{q}},\frac{\sqrt{p/q}}{\sqrt{q}}, \frac{\left(\sqrt{p/q}\right)^{2}}{\sqrt{q}}, \ldots ,\frac{\left(\sqrt{p/q}\right)^{n-1}}{\sqrt{q}},\left(\sqrt{p/q}\right)^{n}\right],\\
{}^T \tilde{\mathbf{v}}_{m}&=C_{m}\cdot \left[1,\frac{\lambda _{m}}{\sqrt{q}},\left(\frac{\lambda _{m}\tilde{U}_{1}^{(m)}}{\sqrt{q}}-\frac{\tilde{U}_{0}^{(m)}}{\sqrt{p}}\right),\ldots ,\left(\frac{\lambda _{m}\tilde{U}_{n-1}^{(m)}}{\sqrt{q}}-\frac{\tilde{U}_{n-2}^{(m)}}{\sqrt{p}}\right),\sqrt{q}\left(\frac{\lambda _{m}\tilde{U}_{n}^{(m)}}{\sqrt{q}}-\frac{\tilde{U}_{n-1}^{(m)}}{\sqrt{p}}\right)\right],\\
&\quad \text{where}\ C_{m}=\frac{1}{\sqrt{\frac{n}{2p}\frac{\sin^{2}\varphi _{m}}{\sin^{2}\theta _{m}}}},\ \tilde{U}_{j}^{(m)}=\frac{\sin(j+1)\theta _{m}}{\sin\theta _{m}}\ \text{and}\ \cos \varphi _{m}=\lambda _{m},\ \text{for}\ 1\leq m\leq n,\\
{}^T \tilde{\mathbf{v}}_{n+1}&=\sqrt{\frac{p-q}{\left(p/q\right)^{n+1}-2q}}\cdot \left[1,\frac{-1}{\sqrt{q}},\frac{\sqrt{p/q}}{\sqrt{q}} ,\frac{-\left(\sqrt{p/q}\right)^{2}}{\sqrt{q}}, \ldots ,\frac{(-1)^{n}\left(\sqrt{p/q}\right)^{n-1}}{\sqrt{q}},(-1)^{n+1}\left(\sqrt{p/q}\right)^{n}\right].
\end{align*}
\end{enumerate}
\end{lem}
The eigenvalues $1=\lambda_{0}$ and $-1=\lambda_{n+1}$ are distinguished from the other eigenvalues $\lambda_{j}\ (j=1,\ldots ,n)$, that is, the
eigenvalues $\lambda_{j}\ (j=1,\ldots ,n)$ are continuously distributed while 
the eigenvalues $1$ and $-1$ remain as mass points when $n\to \infty $. This structure of the eigenspace affects the weak limit theorem. 
\newline
\newline 
{\bf Proof of Lemma \ref{eigenJn}.}

Although it is well-known result (see e.g.\ \cite{VanMieghem2011}) but we show the proof because of improvement of the readability. Let $D_{k}=\det (\lambda I_{k}-J_{k})$ and $E_{k}=\det (\lambda I_{k}-\sqrt{pq}A_{k})=(\sqrt{pq})^{k}\det (\lambda /\sqrt{pq}I_{k}-A_{k})$, where $A_{k}$ is the adjacency matrix of the path $P_{k}$. We can easily obtain the following recurrence equation:
\begin{eqnarray}\label{eqdk}
D_{k+2}=\lambda ^{2}E_{k}-\lambda E_{k-1}+pqE_{k-2}.
\end{eqnarray}
On the other hand, we have the following recurrence equation by a simple calculation:
\begin{align*}
E_{0}/(\sqrt{pq})^{0}&=1,\\
E_{1}/(\sqrt{pq})^{1}&=\lambda /\sqrt{pq},\\
\left(\lambda /\sqrt{pq}\right)E_{k}/(\sqrt{pq})^{k}&=E_{k+1}/(\sqrt{pq})^{k+1}+E_{k-1}/(\sqrt{pq})^{k-1}\quad (k=2,3,\ldots ).
\end{align*}
Noting that $|\lambda |\leq 1$ by the Perron-Frobenius theorem since $0<p,q<1$ and $p+q=1$, this implies 
\begin{align}\label{eqek}
E_{k}=
\begin{cases}
\lambda ^{k}(p^{k+1}-q^{k+1})/(p-q)& \text{if $|\lambda |=1$},\\
(\sqrt{pq})^{k}\tilde{U}_{k}(\lambda /\sqrt{pq}) & \text{if $|\lambda |<1$},
\end{cases}
\end{align}
where $\tilde{U}_{k}(x)$ is the (monic) Chebyshev polynomial of the second kind, i.e., $\tilde{U}_{k}(x)=\sin (k+1)\theta /\sin \theta $ with $x=2\cos \theta $. Combining Eq.\ (\ref{eqdk}) with Eq.\ (\ref{eqek}), we have $D_{k+2}=0$ for $|\lambda |=1$. For $|\lambda |<1$, we have 

\begin{eqnarray*}
D_{n+2}=(\sqrt{pq})^{n}(\lambda ^{2}-1)\frac{\sin (n+1)\theta }{\sin \theta },\ \ \text{with}\ \cos \theta =\lambda /(2\sqrt{pq}). 
\end{eqnarray*}
In this case, the solutions of $D_{n+2}=0$ are $\theta =m\pi/(n+1)\ (m=1,\ldots ,n)$. Therefore we obtain the desired result. 

Next we estimate the eigenvector. Let $P_{0}(\lambda )=1$, $P_{1}(\lambda )=\lambda $, $P_{2}(\lambda )=\det(\lambda I_{2}-\sqrt{q}A_{2})=\lambda P_{1}(\lambda )-qP_{0}(\lambda )$, 
\begin{eqnarray*}
P_{k}(\lambda )=\det 
\begin{bmatrix}
\lambda & -\sqrt{q} & & & & \\
-\sqrt{q} & \lambda & -\sqrt{pq} & & \mbox{\smash{\huge\textit{O}}} & \\
 & -\sqrt{pq} & \ddots & \ddots & & \\
 & & \ddots & \ddots & -\sqrt{pq} \\
\mbox{\smash{\huge\textit{O}}} & & & -\sqrt{pq} & \lambda \\
\end{bmatrix}\ (2\leq k\leq n+1), 
\end{eqnarray*}
where the size of the matrix defining $P_{k}(\lambda )$ is $k\times k$. Then we have the eigenvector $\tilde{\mathbf{v}}_{m}=\mathbf{v}_{m}/\lVert\mathbf{v}_{m}\rVert$ (see Lemma 1.91 of \cite{HoraObata2007}), where
\begin{eqnarray*}
{}^{T}\mathbf{v}_{m}=
\left[
P_{0}(\lambda _{m}), P_{1}(\lambda _{m})/\sqrt{q}, P_{2}(\lambda _{m})/\sqrt{q(pq)},\ldots , P_{n+1}(\lambda _{m})/\sqrt{q(pq)^{n}}
\right].
\end{eqnarray*}
For $3\leq k\leq n+1$, we have $P_{k}(\lambda _{m})=\lambda _{m}E_{k-1}-qE_{k-2}(\lambda _{m})$. Combining this with Eq.\ (\ref{eqek}), we obtain 
\begin{align*}
P_{k}(\lambda _{m})
=
\begin{cases}
p^{k-1} & \text{if $m=0$},\\
\lambda _{m}(\sqrt{pq})^{k-1}\tilde{U}_{k-1}\left(\lambda _{m}/\sqrt{pq}\right)-q(\sqrt{pq})^{k-2}\tilde{U}_{k-2}\left(\lambda _{m}/\sqrt{pq}\right) & \text{if $1\leq m\leq n$},\\
-(-p)^{k-1} & \text{if $m=n+1$}.
\end{cases}
\end{align*}

Finally, we calculate $\lVert\mathbf{v}_{m}\rVert^{2}$. By a direct calculation, we have 
\begin{align*}
\lVert\mathbf{v}_{\pm 1}\rVert^{2}=\frac{(p/q)^{n+1}-2q}{p-q}.
\end{align*}
Let $\tilde{U}_{k}^{(m)}\equiv \tilde{U}_{k}(\lambda _{m}/\sqrt{pq})=\sin (k+1)\theta _{m}/\sin \theta_{m}$ with $\theta _{m}=m\pi/(n+1)$ for $1\leq m\leq n$ and $0\leq k\leq n$. Noting that $\tilde{U}_{-1}^{(m)}=\tilde{U}_{n}^{(m)}=0$, we obtain 
\begin{align*}
\lVert\mathbf{v}_{m}\rVert^{2}
&=
1+\sum _{j=0}^{n-1}\left(\frac{\lambda _{m}\tilde{U}_{j}^{(m)}}{\sqrt{q}}-\frac{\tilde{U}_{j-1}^{(m)}}{\sqrt{p}}\right)^{2}
+
q\left(\frac{\lambda _{m}\tilde{U}_{n}^{(m)}}{\sqrt{q}}-\frac{\tilde{U}_{n-1}^{(m)}}{\sqrt{p}}\right)^{2}\\
&=
1+\frac{p\lambda _{m}^{2}+q}{pq}\sum _{j=0}^{n-1}\left(\tilde{U}_{j}^{(m)}\right)^{2}
-\frac{2\lambda _{m}^{2}}{\sqrt{pq}}\sum _{j=0}^{n-1}\tilde{U}_{j}^{(m)}\tilde{U}_{j-1}^{(m)}
-\left(\tilde{U}_{n-1}^{(m)}\right)^{2}
.
\end{align*}
Using $(\lambda _{m}/\sqrt{pq})\tilde{U}_{j}^{(m)}=\tilde{U}_{j+1}^{(m)}+\tilde{U}_{j-1}^{(m)}$ and $\tilde{U}_{n-1}^{(m)}=(-1)^{m}$, we have 
\begin{align*}
\lVert\mathbf{v}_{m}\rVert^{2}
=
1+\frac{\sin^{2}\varphi_{m}}{p}\sum _{j=0}^{n-1}\left(\tilde{U}_{j}^{(m)}\right)^{2}-\left(\tilde{U}_{n-1}^{(m)}\right)^{2}
=
\frac{\sin^{2}\varphi_{m}}{p}\sum _{j=0}^{n-1}\left(\tilde{U}_{j}^{(m)}\right)^{2},
\end{align*}
with $\sin \varphi _{m}=\lambda _{m}$. On the other hand, 
\begin{align*}
\sum _{j=0}^{n-1}\left(\tilde{U}_{j}^{(m)}\right)^{2}
&=
\sum _{j=0}^{n-1}\frac{\sin^{2}(j+1)\theta _{m}}{\sin^{2}\theta _{m}}
=
\frac{1}{2\sin ^{2}\theta _{m}}\left\{n-\sum _{j=0}^{n}\cos 2j\theta _{m}\right\}
=
\frac{1}{2\sin ^{2}\theta _{m}}\left\{n-\tilde{U}_{n}^{(m)}\cos n\theta _{m}\right\}\\
&=
\frac{n}{2\sin ^{2}\theta _{m}}.
\end{align*}
This completes the proof. 
\qed

Next we obtain the eigenspace of the time evolution operator $U^{(n)}$. Note that there are $2n+2$ numbers of eigenvalues by the definition of the Hilbert space defining our DTQW.
\begin{lem}\label{eigenQW}
\begin{enumerate}
\item 
Let $\mu _{k}\ (k=0,\pm 1,\pm 2,\ldots, \pm n,n+1)$ be the eigenvalues of the time evolution operator $U^{(n)}$. Then we have 
\begin{align*}
\mu _{0}&=1,\\
\mu _{\pm m}&=e^{\pm i\varphi_{m}}\ (m=1,2,\ldots ,n),\\
\mu _{n+1}&=-1,
\end{align*}
where $i=\sqrt{-1}$ and $\cos \varphi _{m}=\lambda _{m} \in \mathrm{Spec}\ (J_{n+2})\setminus \{1,-1\}$ with $\varphi _{m}\in (0,\pi)$. 
\item 
Let $\mathbf{u}_{k}$ be the eigenvector corresponding to an eigenvalue $\mu _{k}$ for each $k=0,\pm 1,\pm 2,\ldots, \pm n,n+1$. Then we have 
\begin{eqnarray*}
\mathbf{u}_{k}=\sum _{j=0}^{n+1}\ket{j}\otimes \left(u_{j,L}^{(k)}\ket{L}+u_{j,R}^{(k)}\ket{R}\right),
\end{eqnarray*}
with 
\begin{align*}
u_{j,L}^{(k)}
&=
\begin{cases}
\sqrt{\frac{p-q}{\left(p/q\right)^{n+1}-2q}}\cdot \left(\mu _{k}\sqrt{p/q}\right)^{j-1} & \text{if\ $1\leq j\leq n+1,\ k=0, n+1$},\\
-i\cdot \left(\tilde{U}_{j-1}^{(k)}-\mu _{k}\sqrt{q/p}\ \tilde{U}_{j-2}^{(k)}\right)\Big/\sqrt{\frac{n}{p}\frac{\sin^{2} \varphi_{k}}{\sin ^{2}\theta _{k}}} & \text{if\ $1\leq j\leq n+1,\ k=\pm 1,\pm 2,\ldots, \pm n$},\\
0 & \text{otherwise},
\end{cases}
\\
u_{j,R}^{(k)}
&=
\begin{cases}
\mu _{k}\sqrt{\frac{p-q}{\left(p/q\right)^{n+1}-2q}}\cdot \left(\mu _{k}\sqrt{p/q}\right)^{j} & \text{if\ $0\leq j\leq n,\ k=0, n+1$},\\
-i\cdot \left(\mu _{k}\tilde{U}_{j}^{(k)}-\sqrt{q/p}\ \tilde{U}_{j-1}^{(k)}\right)\Big/\sqrt{\frac{n}{p}\frac{\sin^{2} \varphi_{k}}{\sin ^{2}\theta _{k}}} & \text{if\ $0\leq j\leq n,\ k=\pm 1,\pm 2,\ldots, \pm n$},\\
0 & \text{otherwise}.
\end{cases}
\end{align*}
\end{enumerate}
\end{lem}
The eigenvalues of the Jacobi matrix $J_{n+2}$ are projections of the eigenvalues of $U^{(n)}$ 
to the real axis from the unit circle of the complex plane. Therefore the eigenvalues (of $U^{(n)}$) $1=\mu_{0}$ and 
$-1=\mu_{n+1}$ are also distinguished from other eigenvalues (of $U^{(n)}$)
$\mu_{\pm j}\ (j=1,\ldots ,n)$ because the eigenvalues (of $J_{n+2}$) $1=\lambda_{0}$ and $-1=\lambda_{n+1}$ are distinguished from the other eigenvalues (of $J_{n+2}$) $\lambda_{j}\ (j=1,\ldots ,n)$. The eigenvectors corresponding to the eigenvalues 
$1$ and $-1$ give the mass at the origin in Theorem \ref{thmtimeavedist}. 
\newline
\newline
{\bf Proof of Lemma \ref{eigenQW}.}

Let $\tilde{\mathbf{v}}_{m}(j)\ (j=0,1,\ldots ,n,n+1)$ be the $j$-th component of the eigenvector $\tilde{\mathbf{v}}_{m}$ corresponding to $\lambda _{m}\in \mathrm{Spec}\ (J_{n+2})$. We define the following vectors:
\begin{align*}
\mathbf{a}_{m}
&=\tilde{\mathbf{v}}_{m}(0)\ket{0,R}+\sum _{j=1}^{n}\tilde{\mathbf{v}}_{m}(j)\ket{j}\otimes \left(\sqrt{q}\ket{L}+\sqrt{p}\ket{R}\right)+\tilde{\mathbf{v}}_{m}(n+1)\ket{n+1,L},\\
\mathbf{b}_{m}
&=S^{(n)}\mathbf{a}_{m}.
\end{align*}
It is easy to see that $U^{(n)}\mathbf{a}_{m}=\mathbf{b}_{m}$ and $U^{(n)}\mathbf{b}_{m}=2\lambda _{m}\mathbf{b}_{m}-\mathbf{a}_{m}$. This implies that $U^{(n)}$ is a linear transformation on the Hilbert space spanned by the two vectors $\mathbf{a}_{m}$ and $\mathbf{b}_{m}$. Moreover, the inner product $(\mathbf{a}_{m},\mathbf{b}_{m})$ equals $\lambda _{m}\lVert\mathbf{a}_{m}\rVert\lVert\mathbf{b}_{m}\rVert=\lambda _{m}$. Therefore, for $m=0$ (resp. $m=n+1$), i.e., $\lambda _{m}=1$ (resp. $\lambda _{m}=-1$), case, we have $\mathbf{b}_{m}=\lambda _{m}\mathbf{a}_{m}$. Thus $\lambda _{0}=1$ and $\lambda _{n+1}=-1$ are eigenvalues of $U^{(n)}$ with corresponding eigenvectors $\mathbf{a}_{0}$ and $\mathbf{a}_{n+1}$, respectively. For $m\neq 0,n+1$ ($\lambda _{m}\neq \pm 1$) case, $\{\mathbf{a}_{m},\mathbf{b}_{m}\}$ is a base of the Hilbert space spanned by $\mathbf{a}_{m}$ and $\mathbf{b}_{m}$. The representation matrix of $U^{(n)}$ with respect to the base $\{\mathbf{a}_{m},\mathbf{b}_{m}\}$ is 
$\left[
\begin{smallmatrix}
0 & -1 \\
1 & 2\lambda _{m}
\end{smallmatrix}
\right]$. 
Therefore, the eigenvalues are $e^{\pm i\varphi_{m}}$ and the corresponding eigenvectors are $\mathbf{a}_{m}-e^{\pm i\varphi_{m}}\mathbf{b}_{m}$. It is easy to see that $\lVert \mathbf{a}_{m}-e^{\pm i\varphi_{m}}\mathbf{b}_{m}\rVert^{2}=2(1-\lambda _{m}^{2})=2\sin^{2}\varphi _{m}$. Combining these facts and Lemma \ref{eigenJn}, we have the explicit form of the eigenvectors after a long but simple calculation. 
\qed

Now we estimate the distribution $\overline{p}_{i}^{(n)}$ of the random variable $\overline{X}_{i}^{(n)}$. By the assumption of the choice of the initial state, we have 
\begin{eqnarray*}
\overline{p}_{i}^{(n)}(x)=
\begin{cases}
\lim _{T\to \infty }\frac{1}{T}\sum_{t=0}^{T-1}
\left\lVert \left(\bra{x}\otimes I_{2}\right)\left(U^{(n)}\right)^{t}(\ket{0}\otimes \ket{R})\right\rVert^{2}
&\text{if $i=0$,}\\
\lim _{T\to \infty }\frac{1}{T}\sum_{t=0}^{T-1}
\left\{
\frac{1}{2}
\sum_{J=L,R}
\left\lVert \left(\bra{x}\otimes I_{2}\right)\left(U^{(n)}\right)^{t}(\ket{i}\otimes \ket{J})\right\rVert^{2}
\right\}
&\text{if $1\leq i\leq n$,}\\
\lim _{T\to \infty }\frac{1}{T}\sum_{t=0}^{T-1}
\left\lVert \left(\bra{x}\otimes I_{2}\right)\left(U^{(n)}\right)^{t}(\ket{n+1}\otimes \ket{L})\right\rVert^{2}
&\text{if $i=n+1$.}
\end{cases}
\end{eqnarray*}
Using the spectral decomposition $\left(U^{(n)}\right)^{t}=\sum _{k}\mu _{k}^{t}\mathbf{u}_{k}\mathbf{u}_{k}^{\dag}$ and $\lim _{T\to \infty }(1/T)\sum_{t=0}^{T-1}e^{i\theta t}=\delta _{0}(\theta )\ (\text{mod} \ 2\pi)$, we obtain 
\begin{eqnarray*}
\overline{p}_{i}^{(n)}(x)=
\begin{cases}
\sum _{k}\left\{(|u_{x,L}^{(k)}|^{2}+|u_{x,R}^{(k)}|^{2})\times |u_{0,R}^{(k)}|^{2}\right\}
&\text{if $i=0$,}\\
\frac{1}{2}
\sum_{J=L,R}\left[
\sum _{k}\left\{(|u_{x,L}^{(k)}|^{2}+|u_{x,R}^{(k)}|^{2})\times |u_{i,J}^{(k)}|^{2}\right\}
\right]
&\text{if $1\leq i\leq n$,}\\
\sum _{k}\left\{(|u_{x,L}^{(k)}|^{2}+|u_{x,R}^{(k)}|^{2})\times |u_{n+1,L}^{(k)}|^{2}\right\}
&\text{if $i=n+1$,}
\end{cases}
\end{eqnarray*}
because all eigenvalues of $U^{(n)}$ are nondegenerate. Note that it is enough to treat the first two cases of $\overline{p}_{i}^{(n)}(x)$ in our setting. From Lemma \ref{eigenQW}, we have the following expressions:
\begin{align}\label{eqp0}
\overline{p}_{0}^{(n)}(j)&=
\frac{2(p-q)^{2}}{\left\{(p/q)^{n+1}-2q\right\}^{2}}\left\{\delta _{0}(j)+\left(1-\delta _{0}(j)\right)\frac{1}{q}(p/q)^{j-1}\right\}\notag \\
&+\frac{2p}{n^{2}}\sum_{m=1}^{n}\left(\frac{\sin\theta_{m}}{\sin\varphi_{m}}\right)^{4}\left\{p\left(\tilde{U}_{j}^{(m)}\right)^{2}-\cos2\varphi_{m}\left(\tilde{U}_{j-1}^{(m)}\right)^{2}+q\left(\tilde{U}_{j-2}^{(m)}\right)^{2}\right\},
\end{align}
for $i=0$ and 
\begin{align}\label{eqpi}
\overline{p}_{i}^{(n)}(j)&=
\frac{(p-q)^{2}}{q\left\{(p/q)^{n+1}-2q\right\}^{2}}(p/q)^{i-1}\left\{\delta _{0}(j)+\left(1-\delta _{0}(j)\right)\frac{1}{q}(p/q)^{j-1}\right\}\notag \\
&+\frac{1}{n^{2}}\sum_{m=1}^{n}\left(\frac{\sin\theta_{m}}{\sin\varphi_{m}}\right)^{4}
\left\{p\left(\tilde{U}_{j}^{(m)}\right)^{2}-\cos2\varphi_{m}\left(\tilde{U}_{j-1}^{(m)}\right)^{2}+q\left(\tilde{U}_{j-2}^{(m)}\right)^{2}\right\}\notag \\
&\quad \quad \quad \quad \quad \quad \quad \quad \quad \quad \quad \quad \times 
\left\{p\left(\tilde{U}_{i}^{(m)}\right)^{2}-\cos2\varphi_{m}\left(\tilde{U}_{i-1}^{(m)}\right)^{2}+q\left(\tilde{U}_{i-2}^{(m)}\right)^{2}\right\},
\end{align}
for $1\leq i\leq n$. 

From now on, we estimate $\PM(\overline{X}_{i}^{(n)}\leq an)=\sum_{j=0}^{\lfloor an \rfloor}\overline{p}_{i}^{(n)}(j)\ (0\leq a\leq 1)$ where $\lfloor z \rfloor$ denotes the largest integer not greater than $z$. At first, we consider the first terms in Eqs.\ (\ref{eqp0}) and (\ref{eqpi}). We can easily see that
\begin{eqnarray*}
\sum_{j=0}^{\lfloor an \rfloor}\left\{\delta _{0}(j)+\left(1-\delta _{0}(j)\right)\frac{1}{q}(p/q)^{j-1}\right\}
=
1+\frac{1}{q}\times \frac{1-(p/q)^{\lfloor an \rfloor}}{1-(p/q)}.
\end{eqnarray*}
Therefore the first terms converge to zero if $p\geq q$ and finite $i$ case and infinite $i$ case. For $p<q$ and finite $i$ case, the first term in Eq.\ (\ref{eqp0}) converges to $1-p/q$, also that of Eq.\ (\ref{eqpi}) converges to $(1-p/q)(p/q)^{i-1}/(2q)$. 

Next we estimate the second term in Eq.\ (\ref{eqp0}). In this case we should consider the following summation:
\begin{eqnarray*}
\frac{1}{n^{2}}\sum_{j=0}^{\lfloor an \rfloor}\sum_{m=1}^{n}\left(\frac{\sin\theta_{m}}{\sin\varphi_{m}}\right)^{4}\left\{p\left(\tilde{U}_{j}^{(m)}\right)^{2}-\cos2\varphi_{m}\left(\tilde{U}_{j-1}^{(m)}\right)^{2}+q\left(\tilde{U}_{j-2}^{(m)}\right)^{2}\right\}.
\end{eqnarray*}
Note that this summation is finite thus we can exchange the order. Using the following relation:
\begin{eqnarray*}
\sum_{j=0}^{l}\left(\tilde{U}_{j}^{(m)}\right)^{2}=\frac{1}{2\sin^{2}\theta_{m}}\left\{l-\cos(l\theta_{m})\tilde{U}_{l}^{(m)}\right\}, 
\end{eqnarray*}
we obtain
\begin{align*}
& \frac{1}{n^{2}}\sum_{m=1}^{n}\sum_{j=0}^{\lfloor an \rfloor}\left(\frac{\sin\theta_{m}}{\sin\varphi_{m}}\right)^{4}\left\{p\left(\tilde{U}_{j}^{(m)}\right)^{2}-\cos2\varphi_{m}\left(\tilde{U}_{j-1}^{(m)}\right)^{2}+q\left(\tilde{U}_{j-2}^{(m)}\right)^{2}\right\}\\
\sim &
\frac{1}{n^{2}}\sum_{m=1}^{n}\left(\frac{\sin\theta_{m}}{\sin\varphi_{m}}\right)^{4}\left\{1-\cos(2\varphi_{m})\right\}\left\{\lfloor an \rfloor-\cos(\lfloor an \rfloor\theta_{m})\tilde{U}_{\lfloor an \rfloor}^{(m)}\right\}\\
\sim &
a\sum_{m=1}^{n}\left(\frac{\sin\theta_{m}}{\sin\varphi_{m}}\right)^{2}\frac{1}{n}\\
\sim &
a\int_{0}^{\pi}\left(\frac{\sin k}{\sin\varphi(k)}\right)^{2}\frac{dk}{\pi},
\end{align*} 
where $f(n)\sim g(n)$ denotes $\lim_{n\to \infty}f(n)/g(n)=1$ and $\cos \varphi(k)=2\sqrt{pq}\cos k$. This implies the second term in Eq.\ (\ref{eqp0}) converges to the uniform distribution $U(0,1)$ with total mass $2p\int_{0}^{\pi}\left(\sin k/\sin\varphi(k)\right)^{2}dk/\pi$. This total mass can be calculated directly but it is obvious that this integral equals $1-c_{0}$. 

We can estimate the second term in Eq.\ (\ref{eqpi}) by using similar argument as follows:
\begin{align*}
& \frac{1}{n^{2}}\sum_{m=1}^{n}\left(\frac{\sin\theta_{m}}{\sin\varphi_{m}}\right)^{4}
\left\{p\left(\tilde{U}_{j}^{(m)}\right)^{2}-\cos2\varphi_{m}\left(\tilde{U}_{j-1}^{(m)}\right)^{2}+q\left(\tilde{U}_{j-2}^{(m)}\right)^{2}\right\}\\
&\quad \quad \quad \quad \quad \quad \quad \quad \quad \quad \quad \quad \times 
\left\{p\left(\tilde{U}_{i}^{(m)}\right)^{2}-\cos2\varphi_{m}\left(\tilde{U}_{i-1}^{(m)}\right)^{2}+q\left(\tilde{U}_{i-2}^{(m)}\right)^{2}\right\}\\
&\sim a\int_{0}^{\pi}\frac{p\sin\{(i+1)k\}-\cos2\varphi(k)\sin(ik)+q\sin\{(i-1)k\}}{\sin\varphi(k)^{2}}\frac{dk}{\pi}.
\end{align*}
Generally, it is hard to calculate the explicit value of the last integral but it is obvious that this integral equals $1-c_{i}$. This completes the proof. 
\par
\
\par\noindent
{\bf Acknowledgments.} 
One of the authors (N. K.) was supported by the Grant-in-Aid for Scientific Research (C) of Japan Society for the Promotion of Science (Grant No. 21540118). 
Y. I. was supported by the Grant-in-Aid for Young Scientists (B) of Japan Society for the Promotion of Science (Grant No. 23740093).


\begin{small}

\end{small}

\end{document}